\documentclass[reprint,twocolumn,aps,superscriptaddress,preprintnumbers,]{revtex4-1}
\usepackage{physics}
\usepackage{qcircuit}
\usepackage{amsmath}
\usepackage{bm}
\usepackage{graphicx}
\usepackage{mathrsfs}
\usepackage{multirow}
\usepackage[normalem]{ulem}
\usepackage{graphicx}
\usepackage{booktabs, ctable}
\usepackage{xcolor, color, framed}
\usepackage[colorlinks, linkcolor=red,anchorcolor=green,citecolor=blue]{hyperref}
\usepackage{cprotect}

\newcommand{\bos}{\boldsymbol}
\newcommand{\ud}{\mathrm{d}}

\begin{document}

\preprint{RIKEN-iTHEMS-Report-25}

\title{Quantum Simulations of Opinion Dynamics}

\author{Xingyu Guo}
\email{guoxy@m.scnu.edu.cn}
\affiliation{State Key Laboratory of Nuclear Physics and Technology, Institute of Quantum Matter, South China Normal University, Guangzhou 510006, China}
\affiliation{Guangdong Basic Research Center of Excellence for Structure and Fundamental Interactions of Matter, Guangdong Provincial Key Laboratory of Nuclear Science, Guangzhou 510006, China}

\author{Xiaoyang Wang}
\email{xiaoyang.wang@riken.jp}
\affiliation{RIKEN Center for Interdisciplinary Theoretical and Mathematical Sciences (iTHEMS), Wako, Saitama 351-0198, Japan}
\affiliation{RIKEN Center for Computational Science (R-CCS), Kobe 650-0047, Japan}

\author{Lingxiao Wang}
\email{lingxiao.wang@riken.jp}
\affiliation{RIKEN Center for Interdisciplinary Theoretical and Mathematical Sciences (iTHEMS), Wako, Saitama 351-0198, Japan}
\affiliation{Institute for Physics of Intelligence, Graduate School of Science, The University of Tokyo, Bunkyo-ku, Tokyo 113-0033, Japan}

\date{\today}

\begin{abstract}
Consensus formation is a central problem in collective behavior. In this work, we develop quantum models of opinion dynamics that can be exactly solved and implemented on current quantum hardware. By exploiting quantum superposition, measurement-induced state collapse, and entanglement, our framework captures key features of opinion evolution and allows a systematic investigation of how network connectivity shapes consensus formation. We demonstrate our approach using practical quantum circuits and validate representative cases on IBM Quantum devices for the open-chain. These findings pave the way for further exploration into quantum-enhanced social modeling, highlighting the potential of near-term quantum computers for simulating collective behavior in complex systems.
\end{abstract}

% Nature-Summary:
% https://docs.google.com/document/d/1YRUWluHJ3HpgZKWjnxtsJzyd7Ep4uxuorGy8sfUhqj8/edit?usp=sharing
\maketitle

\noindent \emph{Introduction.--} Understanding how opinions form, spread, and stabilize in societies is a crucial problem across physics, sociology, and cognitive science~\cite{Castellano_2009,starnini2025opinion}. Classical opinion dynamics models~\cite{noorazar2020recent,noorazar2020classical}---such as the voter~\cite{holley1975ergodic}, Sznajd~\cite{sznajd2000opinion}, and Deffuant~\cite{deffuant2000mixing} frameworks---have been remarkably successful in capturing collective phenomena through simple interaction rules. However, these approaches remain restricted to probabilistic mixtures of discrete states and face increasing computational challenges when extended to large social networks~\cite{zhang2024opinion,PhysRevLett.124.048301,PhysRevLett.130.037401,peel2022statistical,iacopini2024temporal,avalle2024persistent}. In particular, they cannot account for superposition or non-conventional correlations that may underlie complex decision-making processes~\cite{ozawa2020application,ozawa2021modeling}, which are intrinsic challenges for multi-coupled social networks~\cite{Mikko:2014multilayer}.  

Quantum representations of opinions offer a natural resolution to these limitations. Qubits inherently encode \textit{superposition}, mirroring the coexistence of multiple possibilities prior to individual decisions~\cite{pothos2009quantum}. Quantum systems can also simulate \textit{entanglement}, enabling correlated behaviors and collective dynamics inaccessible to classical probability models~\cite{bruza2015quantum}. Moreover, quantum-inspired methods promise \textit{computational advantages} in exploring large-scale interaction networks where the \textit{curse of dimensionality} emerges, thereby opening new pathways for the scalable study of opinion dynamics.

%%%%%%%%%%%%%%%%%%%%%%%%%%%%%%%%%%%%%%%%%%%%%%%%%%%%%%
\begin{figure*}
    \centering
    \includegraphics[width=0.9\linewidth]{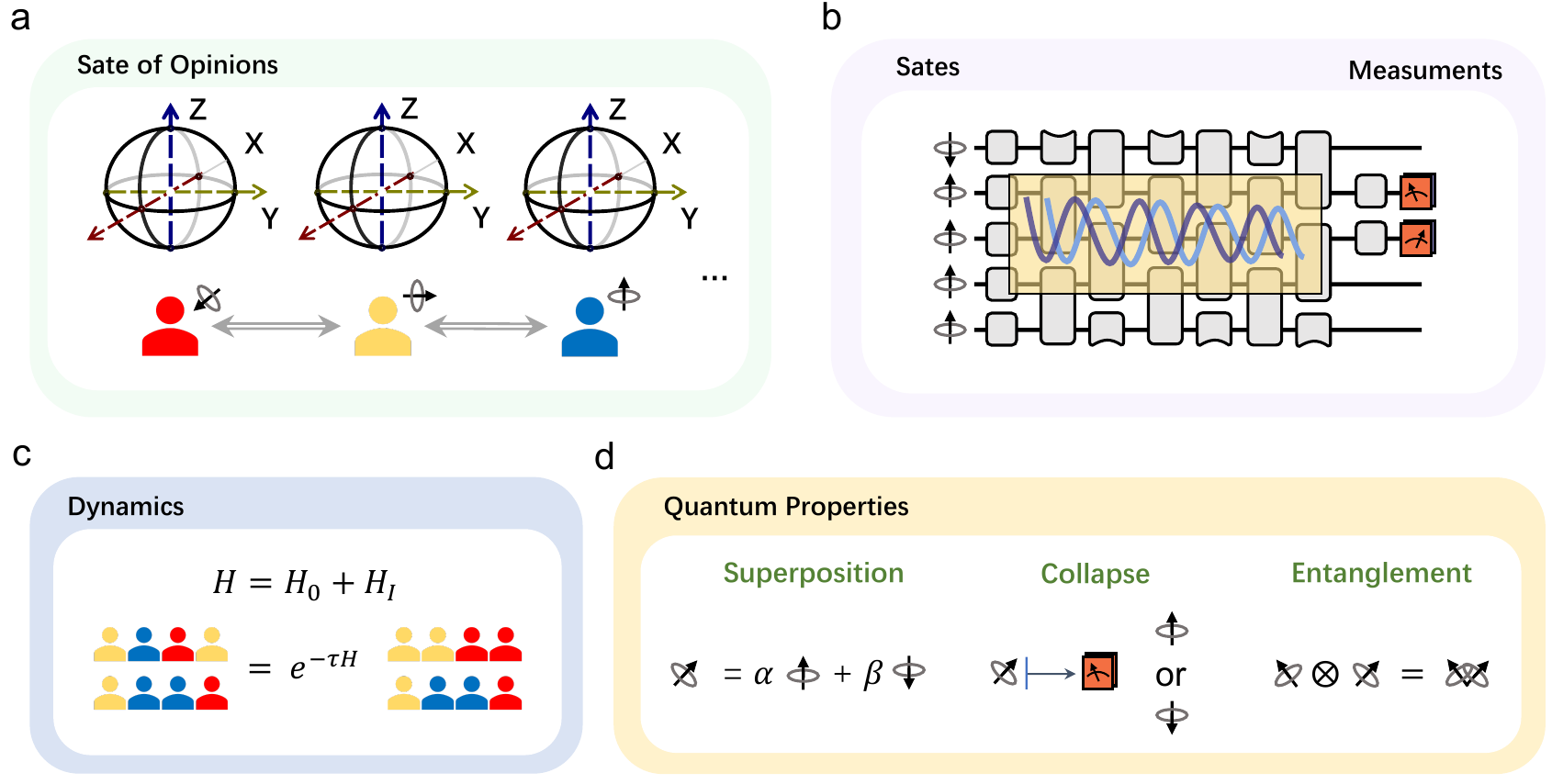}
    \caption{Schematic illustration of the quantum opinion dynamics framework. \textbf{a}. Representation of individual opinions as qubit states on the Bloch sphere. \textbf{b}. Quantum circuit implementation, where initial states evolve under interaction gates and measurements yield observable opinion distributions. \textbf{c}. Dynamics governed by a Hamiltonian $H=H_0+H_I$, with imaginary-time evolution described by $\exp{-\tau H}$. \textbf{d}. Distinct quantum properties relevant to opinion dynamics, including superposition, collapse, and entanglement.}
    \label{fig:flowchart}
\end{figure*}
%%%%%%%%%%%%%%%%%%%%%%%%%%%%%%%%%%%%%%%%%%%%%%%%%%%%%%
Building on these ideas, we establish a fundamental framework for the quantum simulation of opinion dynamics as Fig.~\ref{fig:flowchart} shows. This framework is constructed from two essential components: the specification of \emph{initial beliefs}~\cite{Friedkin2017HowTW,chen2025bayesian}, which encode the starting state of individual opinions, and the modeling of \emph{interactions}~\cite{Castellano_2009}, which govern their persistent influence~\cite{avalle2024persistent}. Together, these elements enable the study of opinion dynamics endowed with fundamental quantum properties, particularly the emergence of steady-state behaviors, metastable consensus, and the real-time dynamical evolution of collective states. Within this setting, we compute observables such as the magnetization, which quantifies opinion alignment, and the entanglement entropy, which characterizes inter-agent correlations in individual opinions. Furthermore, we systematically investigate how different choices of initial beliefs, combined with varying interaction rules, shape the long-term steady-state outcomes of social systems. In addition, two different geometric connections, i.e., round table~\cite{Timpanaro_2009} and leader-follower models~\cite{albi2014boltzmann,dong2017managing}, are discussed to demonstrate the expandability of the quantum framework. This framework opens a novel avenue to investigate social behavior and human decisions, offering a fresh paradigm to study consensus formation, polarization, and emergent collective intelligence in complex societies~\cite{busemeyer2012quantum,starnini2025opinion}.
\\\\
\noindent \emph{Quantum Opinion Dynamics.--} 
We consider a binary opinion denoting pro or con of an agent to an event. In conventional opinion dynamics, the opinion polarization of an agent $i$ can be described by a real number $p_i\in[-1,1]$, with $1$($-1$) denoting an absolute pro(con). In quantum opinion dynamics, we use a qubit to present an agent as Fig.~\ref{fig:flowchart}\textbf{a} shows. The binary opinion of the agent is denoted by $\ket{0}$(pro) and $\ket{1}$(con) states of the qubit as well as the superposition $\ket{\varphi^{(i)}} = \alpha\ket{0}+\beta\ket{1}$. Define the observable $Z_i\equiv \ket{0}\bra{0}-\ket{1}\bra{1}$. $p_i$ is measured on quantum computers by the expectation,
\begin{align}
    p_i = \bra{\varphi^{(i)}}Z_i\ket{\varphi^{(i)}},
\end{align}
where $-1\leq p_i\leq 1$ due to the normalization condition $|\alpha|^2+|\beta|^2=1$ of the quantum state. Thus, the quantum description of a qubit has a direct correspondence to the conventional description of the opinion state of an agent as Fig.~\ref{fig:flowchart} demonstrates.

In conventional, one can simulate opinion dynamics by introducing the scheme of individual responses to their neighborhoods and how opinions exchange among them~\cite{starnini2025opinion}.
% \wxy{Introduce classical simulation of OPD? Describe the necessity of \textit{``agent’s opinion changes in response to its neighbors''} and \textit{``different probabilities of opinion shifts among agents''} given below}
However, compared with the classical simulation of opinion dynamics, the quantum simulation has two fundamental differences:
\begin{itemize}
    \item Quantum states cannot be cloned~\cite{Wootters:1982zz}. Consequently, one cannot simply copy the state of one qubit onto another, making it difficult to model how an agent's opinion changes aligned completely with its neighbors.
    \item Quantum state transitions are inherently symmetric, leading to bidirectional opinion changes. To introduce different probabilities of opinion shifts among agents, additional interaction terms must be incorporated.
\end{itemize}

With these considerations, we introduce a quantum model of opinion dynamics, where the system evolution arises from two fundamental components~\cite{starnini2025opinion}: (i) the \emph{initial belief} of each agent, representing intrinsic predispositions, and (ii) the \emph{interaction} among agents, which drives mutual influence. Rather than explicitly tracking individual opinion states—an typical approach in classical simulations limited by the quantum no-cloning theorem—the interaction term encodes the collective tendency toward consensus, while the initial belief term introduces asymmetry that biases this evolution. Both components are formulated within a Hamiltonian framework, allowing opinion dynamics to be naturally simulated on quantum computing architectures as Fig.~\ref{fig:flowchart}\textbf{b} shows.

The initial belief is described by the Hamiltonian,
\begin{eqnarray}
    H_0 &=& \sum_i a_i \left(Z_i\cos\theta_i + X_i\sin\theta_i\right),
    \label{eq:initial-belief-Hamiltonian}
\end{eqnarray}
where $Z_i$ and $X_i$ are Pauli operators on the $i$th qubit. The ground state of $H_0$ represents the distribution of initial opinions: the parameter $\theta_i$ specifies the orientation of agent $ i$'s initial belief on the Bloch sphere as shown in Fig.~\ref{fig:flowchart}\textbf{a}, while $a_i>0$ quantifies how strongly the initial belief is preserved against its environment. 

The interactions are described by,
\begin{eqnarray}
    H_I &=& \sum_{\langle i, j\rangle} -c_{ij} \left(Z_i Z_{j} + X_i X_{j}\right).
\end{eqnarray}
Here $H_I$ captures the exchange of opinions between neighboring agents $i$ and $j$ in an opinion network, with coupling strength $c_{ij}>0$ controlling the extent to which neighboring agents align.

The full Hamiltonian $H = H_0 + H_I$ is used to govern both dynamical and equilibrium behaviors as Fig.~\ref{fig:flowchart}\textbf{c} shows, and the quantum properties including superposition, measurement-induced collapse and entanglement as shown in Fig.~\ref{fig:flowchart}\textbf{d} can be naturally embedded. Its classical counterpart can be realized as a \textit{Kuramoto-like model}, as Appendix.~\ref{sec:kuramoto} shows.

\begin{figure}[htbp!]
    \centering
    \includegraphics[width=0.9\linewidth]{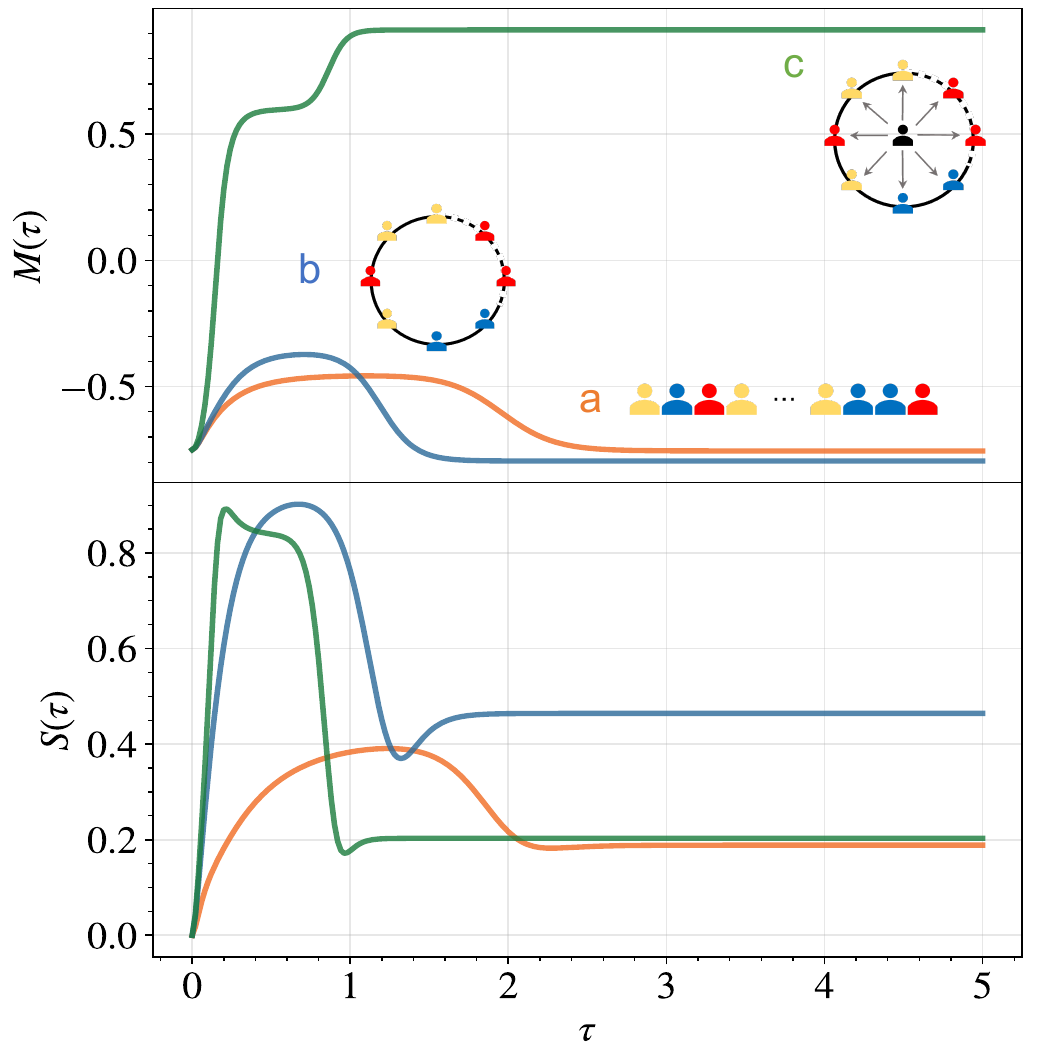}
    \caption{Evolution of magnetization $M(\tau)$, and entanglement entropy $S(\tau)$ between two agent groups. The orange, blue, and green curves correspond to \textbf{a}. one dimensional opinion chain with fixed boundary condition (orange line); \textbf{b}. \textit{round table} connection with periodic boundary condition (blue line); \textbf{c}. \textit{leader-follower} connection (green line).}
    \label{fig:connect-obs}
\end{figure}

As examples, we consider the exchange of opinions in networks shown in Fig.~\ref{fig:connect-obs} \textbf{a(b)}, which is a one dimensional chain with an open(periodic) boundary for agents. Besides, the \textit{leader-follower} model~\cite{albi2014boltzmann,dong2017managing} has an obstinate leader affecting all agents' opinions as shown in Fig.~\ref{fig:connect-obs} \textbf{c}. This \textit{leader} agent can be considered by incorporating $H$ with an effective external field applied to all agents,
\begin{eqnarray}
    H_L&=&\sum_i -d_i Z_i,
\end{eqnarray}
where $d_i>0 (<0)$ controls the strength of the leader influence on the pro(con) opinion of the $i$th agent.

The opinion evolution can be described by either real or imaginary time evolution of $H$. The quantum imaginary-time evolution (QITE) drives the system toward an equilibrium state~\cite{McArdle_19,Motta_20}, corresponding to the collective behavior of consensus formation in the opinion network~\cite{starnini2025opinion}. Real-time unitary evolution, in contrast, reveals dynamical trajectories, oscillatory patterns, and the buildup of entanglement that are absent in classical models, as detailed in Appendix.~\ref{sec:real-time}. For both evolutions, observables such as magnetization, entanglement entropy, and correlation functions provide quantitative probes of consensus, opinion entanglement, and inter-agent dependencies within this unified setting.
\\\\
\noindent \emph{Consensus formation..--} The consensus formation. in opinion dynamics can be demonstrated by QITE,
\begin{eqnarray}
    |\varphi(\tau)\rangle = e^{-\tau H}|\varphi_{0}\rangle/\sqrt{\bra{\varphi_0}e^{-2\tau H}\ket{\varphi_0}} ,
    \label{eq:QITE}
\end{eqnarray}
where $\ket{\varphi_0}$ is an initial opinion state without interaction among agents. Thus, we choose $\ket{\varphi_0}$ to be the ground state of the initial Hamiltonian $H_0$. $\sqrt{\bra{\varphi_0}e^{-2\tau H}\ket{\varphi_0}}$ is a normalization factor ensuring the normalization of the opinion probability distribution. The opinion of each agent is measured by the local observable,
\begin{eqnarray}
     p_{i}(\tau) = \langle \varphi(\tau)|Z_i|\varphi(\tau)\rangle
\end{eqnarray}
after evolving an imaginary time $\tau$. The system is driven to lower-energy configurations as $\tau$ increases. At lower-energy configurations, on the one hand, the interaction Hamiltonian $H_I$ favors alignment among neighboring opinions. On the other hand, the initial Hamiltonian $H_0$ tends to preserve individual predispositions. The observed evolution thus reflects the competition between social conformity and personal resistance: agents with larger $a_i$ values retain their initial opinions more robustly, whereas stronger couplings $c_i$ promote consensus formation.  

%%%%%%%%%%%%%%%%%%%%%%%%%%%%%%%%%%%%%%%%%%%%%%%%%%%%%%
%%%%%%%%%%%%%%%%%%%%%%%%%%%%%%%%%%%%%%%%%%%%%%%%%%%%%%

In our first demonstration, the set-up is $a_i = 1, c_i= 3, \theta_1=(1-10^{-3})\pi, \theta_{i\neq1} = 0$ for both open and periodic chains(which are also named as \textit{round table} hereafter), and $d_i=3$ for the \textit{leader-follower} connection, with eight agents. It means that only one agent will hold the positive opinion initially, then the interactions will potentially change their opinions. If not specified, these parameters are used throughout this letter. Other different initial beliefs and interaction strengths are discussed in Appendix.~\ref{sec:dif-initial}.

The degree of consensus formation is quantified by the magnetization of the opinion distribution, $M(\tau)=\sum_i^N p_i(\tau)$, shown in Fig.~\ref{fig:connect-obs}. This global observable tracks the emergence of consensus or persistent disagreement. At large $\tau$, $M(\tau)$ approaches a plateau, indicating convergence to a stable collective opinion. For the open-chain case, the approach to consensus is marked by a decrease in magnetization around $\tau\sim 2$, while for the round-table case a similar transition occurs around $\tau\sim 1.2$, reflecting the different interaction topologies. By contrast, the \textit{leader-follower} model reaches consensus most rapidly, before $\tau\sim 1$, owing to the strong alignment induced by the leader. 

Different network connectivities also exhibit metastable states before the final consensus plateau. In the open-chain case, a metastable plateau appears for $0.4\lesssim\tau\lesssim1.6$ (upper panel of Fig.~\ref{fig:connect-obs}), where the overall opinion remains relatively positive due to the initial influence of Agent~1. In the round-table case, a similar plateau occurs for $0.4\lesssim\tau\lesssim0.8$, while in the \textit{leader-follower} model it is much shorter, $0.4\lesssim\tau\lesssim0.6$, because the leader drives the system more efficiently toward consensus. Exact diagonalization shows that these metastable states correspond to the first excited state, whereas the final equilibrium corresponds to the ground state\footnote{This behavior is also related to the localization phase transition in many-body physics~\cite{Pal:2010bew}, and directly related to fragmentation in opinion dynamics~\cite{2012PhRvE..85f6117B}.}. Accordingly, the metastable-state energy is higher than that of the final equilibrium.

Another useful quantity is the bipartite entanglement entropy,
$S=-\mathrm{Tr}(\rho_A\log\rho_A)$, where $\rho_A$ is the reduced density matrix of the first $N_s$ agents obtained by tracing out the remaining degrees of freedom~\cite{Calabrese:2005in,Eisert:2008ur,Skinner:2018tjl}. Although the global state remains pure throughout the evolution, $S(\tau)$ measures the buildup of correlations and information exchange across the bipartition. Starting from an initially factorized state, $S(\tau)$ increases as interactions spread correlations through the network. As shown in the lower panel of Fig.~\ref{fig:connect-obs}, the entropy between the first and last four agents exhibits a pronounced peak near the metastable regime, signaling maximal information delocalization and configurational mixing~\cite{Calabrese:2005in,Eisert:2008ur} and serving as a finite-size precursor of the transition toward consensus. Beyond this point, $S(\tau)$ decreases as the system relaxes into a steady state with aligned opinions and saturated correlations. The resulting nonmonotonic profile of $S(\tau)$ therefore provides a clear signature of the imaginary-time transition. A detailed discussion of local correlations is given in Appendix~\ref{sec:local}.

%%%%%%%%%%%%%%%%%%%%%%%%%%%%%%%%%%%%%%%%%%%%%%%%%%%%%%
\begin{figure}[htbp!]
    \centering
    \includegraphics[width=0.85\linewidth]{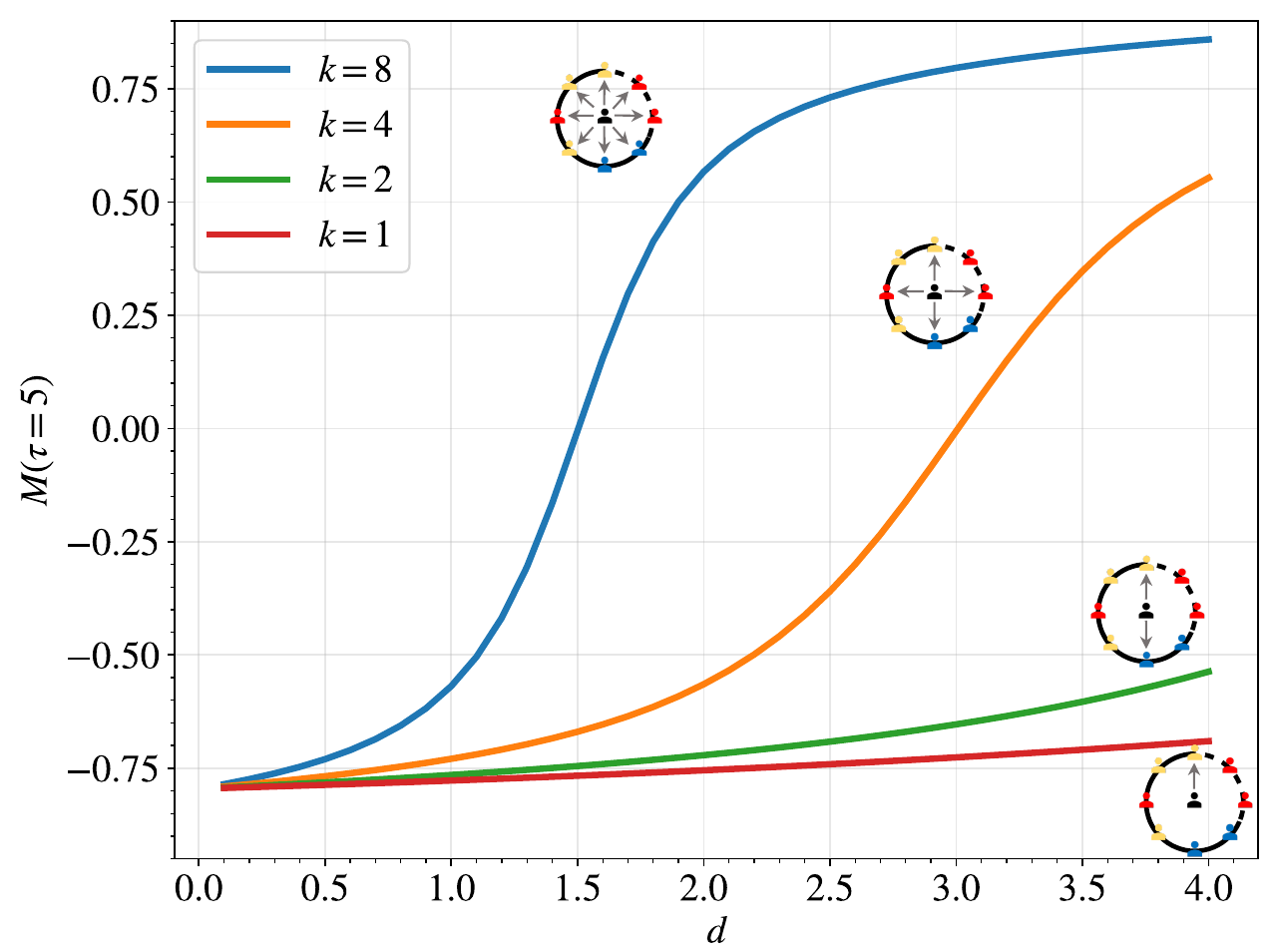}
    \caption{Final magnetization $M(\tau=5)$ as a function of the leader influence strength $d_i$ for different network connectivities $k$. Larger $k$ corresponds to more highly connected networks. Highly connected networks display a sharp transition to a high-magnetization (consensus) state at relatively small $d$, whereas sparsely connected networks require significantly stronger leader influence to achieve comparable alignment. The inset diagrams illustrate representative network topologies for different $k$.}
    \label{fig:leader-mag-final}
\end{figure}
%%%%%%%%%%%%%%%%%%%%%%%%%%%%%%%%%%%%%%%%%%%%%%%%%%%%%%
The \textit{leader–follower} model introduces a leader agent whose opinion exerts a global influence on the rest of the network. To elucidate this effect, we examine the final magnetization as a function of $d$ for networks with different numbers of links $k$, between the leader and followers, as shown in Fig.~\ref{fig:leader-mag-final}. A clear connectivity-dependent transition emerges: in highly connected networks, even a moderate leader influence is sufficient to drive the system into a strongly alignment with the leader, whereas in sparsely connected networks, leader-induced consensus formation is significantly suppressed. This demonstrates that network connectivity qualitatively alters the consensus-formation mechanism, effectively shifting the critical leader strength required for global alignment.

This behavior closely parallels ferromagnetic phase transitions in many-body systems~\cite{2002PhyA..310..260A}, where increased coordination among degrees of freedom lowers the external field required to induce an ordered phase. In this analogy, the leader influence $d$ acts as an external field, while the network connectivity controls the collective susceptibility of the system, jointly determining the emergence of consensus. Details on leader-induced dynamics is given in Appendix~\ref{sec:leader}.

\emph{Hardware implementation.--} To illustrate the applicability of our approach, we implement proof-of-principle hardware demonstration using IBM Quantum platform~\cite{Qiskit}. In this demonstration, qubits encode the opinion states of individual agents, with the initial state $\ket{\varphi_0}$ prepared by single-qubit gates. Then, the imaginary-time dynamics are realized using variational quantum imaginary time evolution (VarQITE)~\cite{McArdle_19,Yuan_2019}, which performs imaginary-time dynamics on a parametrized quantum circuit with fixed depth. Finally, projective measurements in the computational basis yield the opinion distributions and correlations. More details on the hardware implementation are given in Appendix~\ref{app:Hardware implementation details}.
%%%%%%%%%%%%%%%%%%%%%%%%%%%%%%%%%%%%%%%%%%%%%%%%%%%%%%
\begin{figure}[htbp!]
    \centering
    \includegraphics[width=0.95\linewidth]{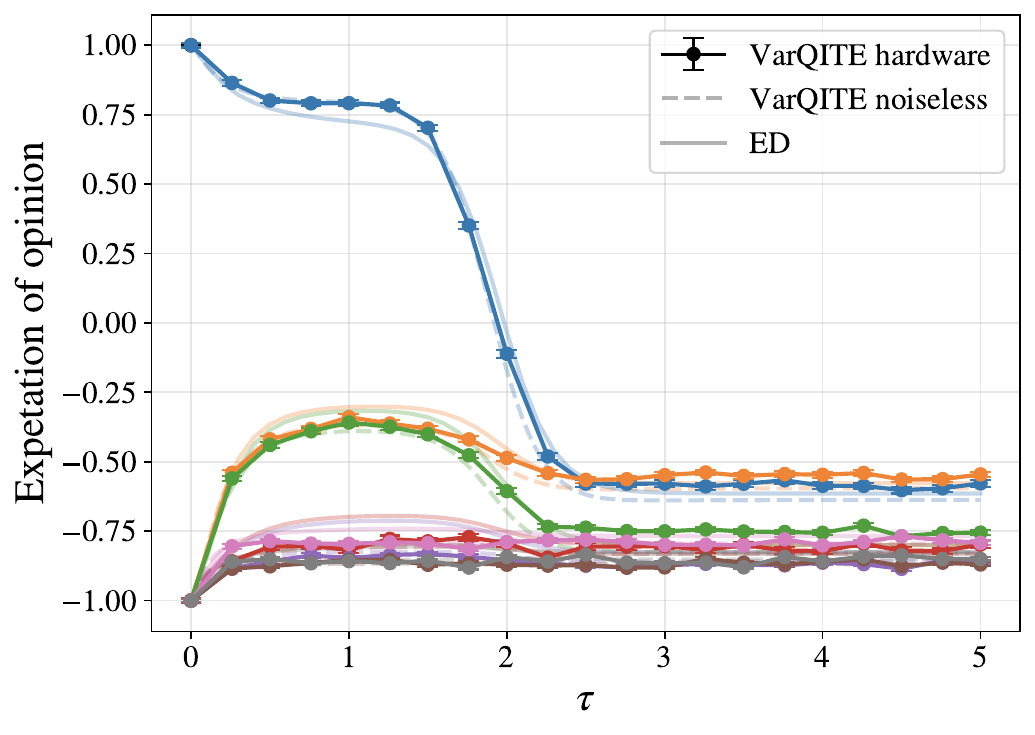}
    \cprotect\caption{Time evolution of the opinion expectations $p_i$ on a one dimensional opinion chain with 8 agents, measured using hardware \verb|ibm_marrakesh|, noiseless circuit simulator, and QITE by the exact diagonalization (ED). The error bar denotes the statistical error from 8192 repeated measurements.}
    \label{fig:op-8-one-ibm}
\end{figure}
%%%%%%%%%%%%%%%%%%%%%%%%%%%%%%%%%%%%%%%%%%%%%%%%%%%%%%
For a one dimensional opinion chain composed of 8 agents, the opinion polarization $p_i$ of each agent $i$ as a function of time $\tau$ is measured on IBM chip \verb|imb_marrakesh| augmented by error mitigation and suppression techniques~\cite{Urbanek_21,Vovrosh_21,Ezzel2023,PhysRevX.11.041039}. In Fig.~\ref{fig:op-8-one-ibm}, the hardware, noiseless VarQITE, and the exact diagonalization(ED) results are denoted by dots, dashed lines, and solid lines, respectively. We see that the opinion dynamics measured on hardware are well consistent with the noiseless VarQITE and ED curves. Specifically, the opinion platforms at time $\tau=0.5\sim 1.25$ and $\tau\geq 2.5$ as well as the opinion transition at $\tau\sim 2$ are observed on quantum chips in the existence of realistic hardware noise. These hardware results validate our theoretical framework and highlight the applicability of quantum opinion dynamics on currently available noisy devices.  

\emph{Summary.--} In this letter, we have established a fundamental framework for quantum simulations of opinion dynamics, leveraging the unique capabilities of quantum computing to capture complex social phenomena. By encoding individual opinions as qubit states and modeling interactions through a Hamiltonian formalism, we have demonstrated how quantum properties such as superposition and entanglement can enrich our understanding of consensus formation and collective behavior. Our results, validated through both exact diagonalization and hardware implementations on IBM Quantum devices, illustrate the potential of quantum simulations to explore nonclassical correlations and dynamical trajectories in social systems.

From a theoretical perspective, our framework also connects naturally to phase-synchronization models such as Kuramoto-type dynamics, and is closely related to Schr\"odinger–Lohe models~\cite{lohe2009non,lohe2010quantum,ha2016collective} on complex-weighted networks~\cite{2024PhRvE.109b4314B}, highlighting a broader link between quantum opinion dynamics and synchronization phenomena. These connections highlight the utility of quantum simulators as a flexible tool for studying collective behavior and open new directions for exploring the interplay between leadership, network structure, and social dynamics in complex networks~\cite{2002RvMP...74...47A,2006PhR...424..175B,2008PhR...469...93A,2020PhR...874....1B}.

\section*{Acknowledgement}
We thank Drs. Tetsuo Hatsuda,  Yohsuke Murase, Riccardo Muolo, Masanao Ozawa, and Isaac Planas-Sitj\`a for helpful discussions.
% Institutions
We thank the DEEP-IN working group at RIKEN-iTHEMS for support in the preparation of this paper.
% Funds
XG is supported by the National Natural Science Foundation of China under Grant No. 12035007.
LW and XW are supported by the RIKEN TRIP initiative (RIKEN Quantum), LX is also supported by JSPS KAKENHI Grant No. 25H01560, and JST-BOOST Grant No.JPMJBY24H9.

\bibliography{refs}

\appendix

\section{Kuramoto-Like Model}
\label{sec:kuramoto}
To obtain the classical correspondence of the quantum Hamiltonian, we treat each spin as a classical vector. Since the original Hamiltonian contains only the $Z$ ($\sigma_z$) and $X$ ($\sigma_x$) Pauli operators, the dynamics are confined to the $X$--$Z$ plane of the Bloch sphere. Consequently, the classical limit reduces to the well-known classical planar rotor-like model.

In the quantum description, $Z_i$ and $X_i$ are Pauli matrices acting on site $i$. In the classical limit, we replace each quantum spin with a unit classical vector $\vec S_i$. Because only the $X$ and $Z$ components appear, the state of agent $i$ can be parametrized by a single angular variable $\phi_i$,
\begin{equation}
    Z_i \;\rightarrow\; \cos\phi_i,\qquad 
    X_i \;\rightarrow\; \sin\phi_i .
\end{equation}
Here $\phi_i$ plays the role of the ``opinion angle'' in the corresponding classical model. Consider the interaction part of the quantum Hamiltonian,
\begin{equation}
    H_I = \sum_{\langle i,j\rangle} -c_{ij}\left( Z_i Z_j + X_i X_j \right).
\end{equation}
Replacing the operators by their classical counterparts yields,
\begin{equation}
    H_I^{\mathrm{cl}}
    = \sum_{\langle i,j\rangle}
      -c_{ij}\bigl( \cos\phi_i \cos\phi_j + \sin\phi_i \sin\phi_j \bigr).
\end{equation}
Using the trigonometric identity 
$\cos(A-B) = \cos A \cos B + \sin A \sin B$, this becomes,
\begin{equation}
    H_I^{\mathrm{cl}}
    = -\sum_{\langle i,j\rangle} c_{ij}\cos(\phi_i - \phi_j).
\end{equation}
For $c_{ij}>0$, minimizing $H_I$ favors $\phi_i \simeq \phi_j$, corresponding to ferromagnetic alignment or, in opinion dynamics, conformity among neighboring agents.

The initial belief part of the Hamiltonian is,
\begin{equation}
    H_0 = \sum_i a_i\bigl( Z_i \cos\theta_i + X_i \sin\theta_i \bigr).
\end{equation}
Under the classical substitution,
\begin{eqnarray}
        H_0^{\mathrm{cl}}
    &=& \sum_i a_i\bigl( \cos\phi_i \cos\theta_i + \sin\phi_i \sin\theta_i \bigr)\nonumber\\
    &=&\sum_i a_i\cos(\phi_i - \theta_i).
\end{eqnarray}
Each agent $i$ experiences a ``bias'' or ``prior opinion'' characterized by angle $\theta_i$ and strength $a_i$.  
Depending on sign conventions, one often writes the classical potential as $-a_i\cos(\phi_i - \theta_i)$ if alignment with $\theta_i$ is energetically favored.

Combining the interaction and field terms, the full classical Hamiltonian reads
\begin{equation}
    H_{\mathrm{cl}}
    = \sum_i a_i \cos(\phi_i - \theta_i) -\sum_{\langle i,j\rangle} c_{ij}\cos(\phi_i - \phi_j).
\end{equation}

If we consider overdamped dynamics, $\dot{\phi}_i \propto -{\partial H_{\mathrm{cl}}}/{\partial \phi_i}$, we can obtain the evolution equation as,
\begin{equation}
    \dot{\phi}_i
    = - a_i \sin(\phi_i - \theta_i) + \sum_{\langle i,j\rangle} c_{ij}\sin(\phi_j - \phi_i).
\end{equation}
The first term pulls each agent toward its intrinsic preferred angle $\theta_i$, while the second term drives social alignment with neighbors. This structure closely parallels the \textbf{Kuramoto model} and related nonlinear phase-oscillator dynamics~\cite{Kuramoto:1975ebm,Acebron:2005zz}. If we redefine the frist term as, $\omega_i \equiv - a_i \sin(\phi_i - \theta_i) $, one can find this is a Kuramoto-like model with time-dependent intrinsic frequencies.
%%%%%%%%%%%%%%%%%%%%%%%%%%%%%%%%%%%%%%%%%%%%%%%%%%%%%%
\begin{figure}[htbp!]
    \centering
    \includegraphics[width=0.9\linewidth]{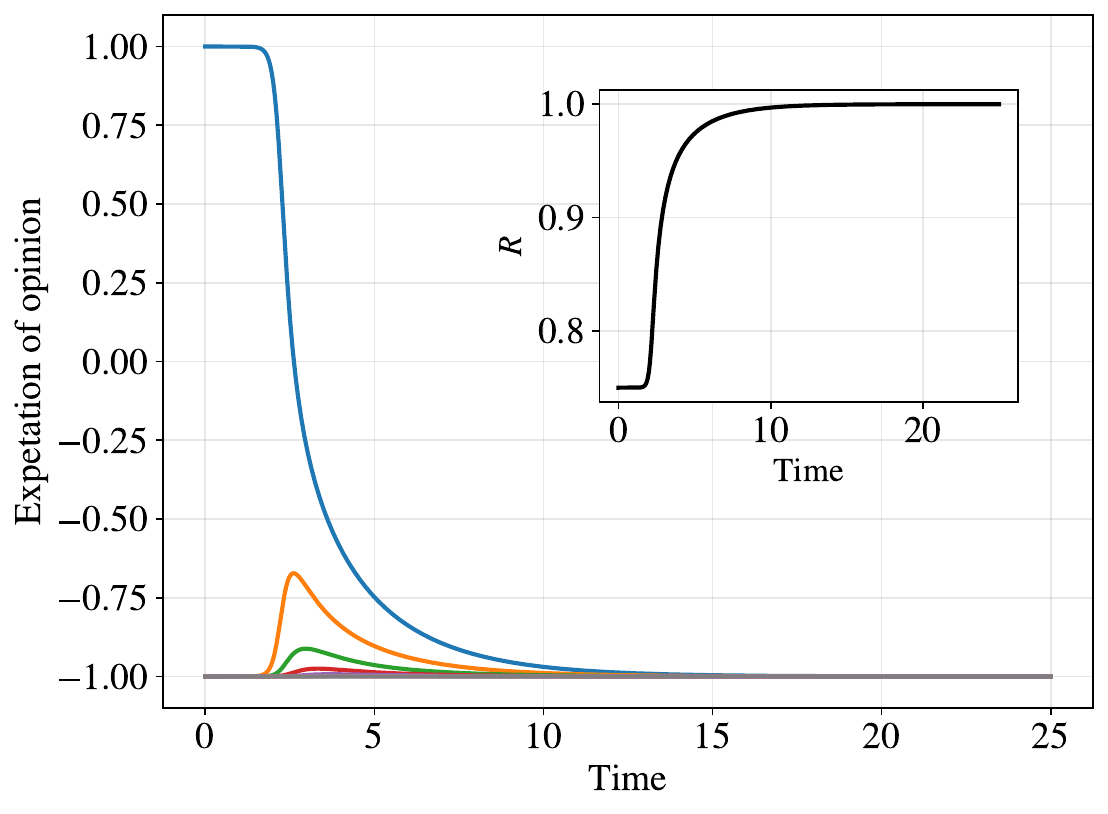}
    \cprotect\caption{Time evolution of the opinion states $\phi_i$ on a one dimensional opinion chain with eight agents, following the Kuramoto-like model. The inset displays the time evolution of the opinion order parameter $R$.}
    \label{fig:kuramoto}
\end{figure}
%%%%%%%%%%%%%%%%%%%%%%%%%%%%%%%%%%%%%%%%%%%%%%%%%%%%%%
As the same set-up in the quantum dynamics, $a_i = 1, c_i= 3, \theta_1=(1-10^{-3})\pi, \theta_{i\neq1} = 0$, we also simulate the classical opinion dynamics by solving the above ordinary differential equations numerically, in which initial opinion states $\phi_i(0)$ following $p_i(\tau=0)$. Fig.~\ref{fig:kuramoto} illustrates the time evolution of the opinion states $\phi_i$ for the classical Kuramoto-like model on an eight-agent opinion chain. The system evolves toward consensus ($\phi_i \approx -1$), but the time required to reach consensus is significantly longer than in the quantum version(around $\tau=2.5$). This delay is evident in the inset, which plots the opinion order parameter $Re^{i\psi} \equiv \frac{1}{N}\sum_{j=1}^{N}e^{i\phi_j}$, the approach to $R=1$ is characterized by a very long "tailing" phase, indicating a much slower relaxation process. Furthermore, this classical dynamics does not exhibit the "stalemate" corresponding to the intermediate meta-stable state observed in the quantum counterpart, highlighting a fundamental difference in the coherence-building mechanisms between the classical and quantum evolution.

\section{Real-Time Evolution}
\label{sec:real-time}

\begin{figure}[h]
    \centering
    \includegraphics[width=1\linewidth]{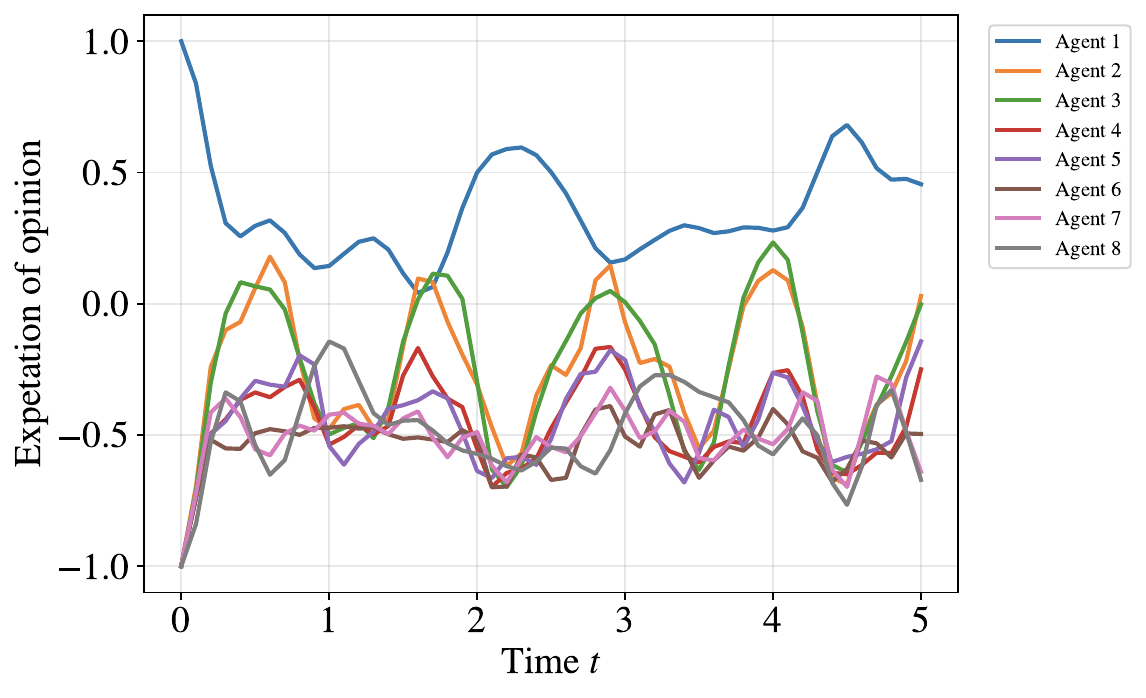}
    \caption{Real-time evolution of opinions for the eight-agent system. The different colored curves correspond to different agents.}
    \label{fig:real-opinions}
\end{figure}

In the real-time evolution, the opinion state at time $t$ is given by
\begin{equation}
        |\varphi(t)\rangle = e^{-i t H}|\varphi_{0}\rangle.
\end{equation}
As emphasized above, the unitary time evolution of quantum systems differs fundamentally from the relaxation-like behavior in classical or imaginary-time dynamics, and thus markedly distinct phenomena can be expected. For direct comparison, we again study the eight-agent system with the same set of parameters used in the imaginary-time case for one dimensional open chain. The time-dependent opinions of individual agents are shown in Fig.~\ref{fig:real-opinions}.  

%%%%%%%%%%%%%%%%%%%%%%%%%%%%%%%%%%%%%%%%%%%%%%%%%%%%%%%
%%%%%%%%%%%%%%%%%%%%%%%%%%%%%%%%%%%%%%%%%%%%%%%%%%%%%%%
In contrast to the monotonic convergence observed under imaginary-time evolution, the real-time dynamics exhibit oscillatory relaxation, reminiscent of a set of coupled quantum oscillators. Despite the oscillations, correlations among agents remain apparent. For instance, Agents~2 and~3, which are nearest neighbors with initially similar opinions, maintain a relatively close alignment throughout both real- and imaginary-time evolutions. This highlights the persistence of local correlations even under qualitatively different dynamical regimes. 

\section{Local observable}
\label{sec:local}
\begin{figure}[htbp!]
    \centering
    \includegraphics[width=1\linewidth]{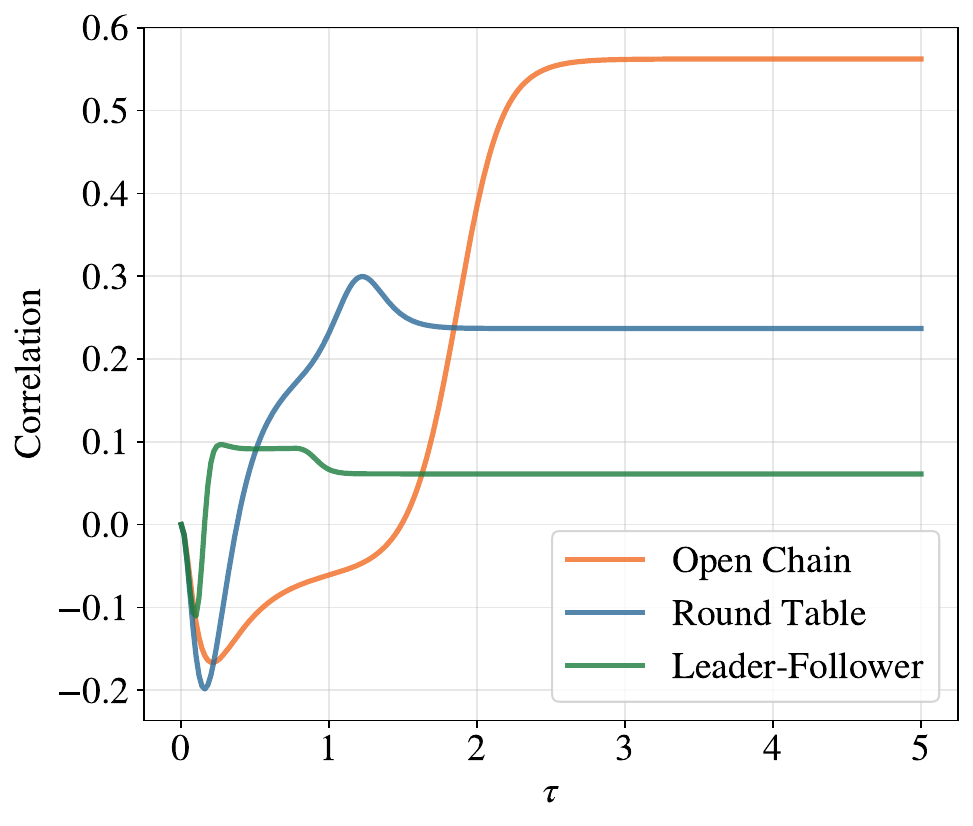}
    \caption{Correlations between agent 1 and 2.}
    \label{fig:correlation}
\end{figure}

Beyond collective observables, it is instructive to probe two-body correlations, particularly between agents that are not direct neighbors. As an illustrative case, Fig.~\ref{fig:correlation} shows the correlation between Agents~1 and~2, defined as,
\begin{equation}
    C_{ij}=\langle Z_i Z_j\rangle - \langle Z_i\rangle \langle Z_j\rangle .
\end{equation}

%%%%%%%%%%%%%%%%%%%%%%%%%%%%%%%%%%%%%%%%%%%%%%%%%%%%%%
%%%%%%%%%%%%%%%%%%%%%%%%%%%%%%%%%%%%%%%%%%%%%%%%%%%%%%
Importantly, this correlation does not merely reflect the similarity of the mean opinions but originates from quantum superpositions of the evolving state. Initially, $C_{ij}=0$, as the Hamiltonian is non-interacting and the system starts from a product state. As time evolves, interactions generate superpositions, and genuine quantum correlations emerge---a feature that has no analogue in classical opinion models.  

\section{Different initial beliefs}
\label{sec:dif-initial}
\begin{figure}[htbp!]
    \centering
    \includegraphics[width=1\linewidth]{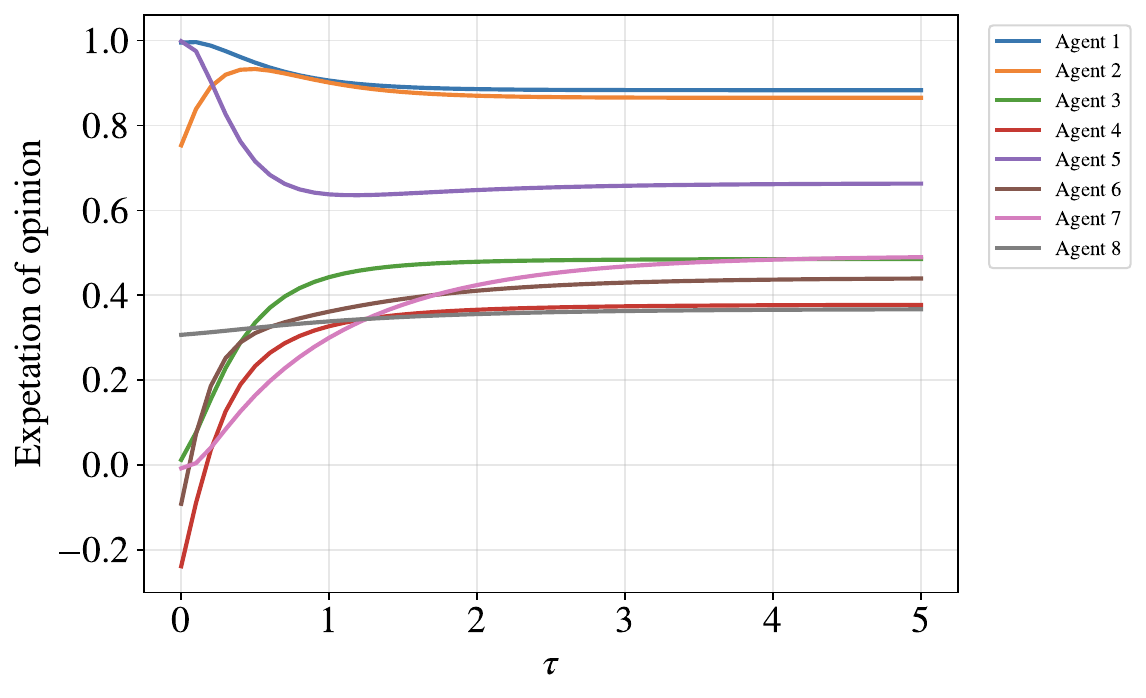}
    \caption{Time evolution of the opinions of eight agents under random initial beliefs and interaction strengths.}
    \label{fig:random}
\end{figure}

To demonstrate the versatility of our framework, we explore different configurations of initial beliefs. First, we consider a system of eight agents with randomly chosen parameters listed in Table~\ref{tab:parameters}. The resulting opinion trajectories, shown in Fig.~\ref{fig:random}, exhibit rich dynamical patterns. Depending on the interplay of initial predispositions and interaction strengths, agents display diverse behaviors including rapid consensus formation, and persistent disagreement.
%%%%%%%%%%%%%%%%%%%%%%%%%%%%%%%%%%%%%%%%%%%%%%%%%%%%%%
\begin{table}[hbtp!]
    \centering
    \resizebox{\columnwidth}{!}
    {\begin{tabular}{|c|c|c|c|c|c|c|c|c|}
    \hline
        & Agent 1 & Agent 2 & Agent 3 & Agent 4 & Agent 5 & Agent 6 & Agent 7 & Agent 8 \\
        \hline
        $a_i$ & 0.068 & 0.831 & 0.573 & 0.598 & 0.220 & 0.432 & 0.637 & 0.371 \\
        $\theta_i / \pi$ & 1.031 & 0.772 & 1.496 & 1.577 & 0.982 & 1.529 & 1.502 & 0.599 \\
        $c_i$ & 0.559 & 0.512 & 0.597 & 0.661 & 0.886 & 0.581 & 0.054 & 0.593 \\
        \hline
    \end{tabular}}
    \caption{Randomly chosen parameters for the system of eight agents.}
    \label{tab:parameters}
\end{table}
%%%%%%%%%%%%%%%%%%%%%%%%%%%%%%%%%%%%%%%%%%%%%%%%%%%%%%

%%%%%%%%%%%%%%%%%%%%%%%%%%%%%%%%%%%%%%%%%%%%%%%%%%%%%%
%%%%%%%%%%%%%%%%%%%%%%%%%%%%%%%%%%%%%%%%%%%%%%%%%%%%%%

%%%%%%%%%%%%%%%%%%%%%%%%%%%%%%%%%%%%%%%%%%%%%%%%%%%%%%
\begin{figure}[htbp!]
    \centering
    \includegraphics[width=1\linewidth]{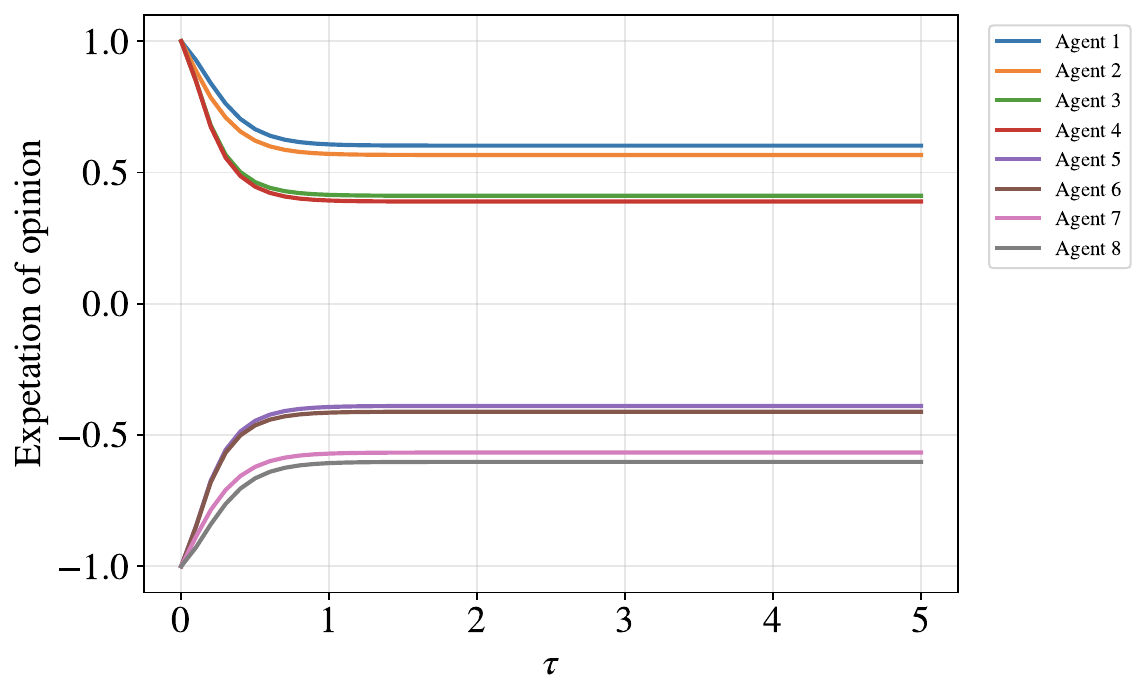}
    \caption{Time evolution of the opinions of eight agents with half positive and half negative initial beliefs.}
    \label{fig:half}
\end{figure}
%%%%%%%%%%%%%%%%%%%%%%%%%%%%%%%%%%%%%%%%%%%%%%%%%%%%%%

As another example, we consider a polarized initial belief configuration where half the agents start with positive opinions and the other half with negative opinions. Specifically, we set $\theta_{\leq4} = 0$ and $\theta_{\geq5} = \pi$, while keeping $a_i = 1$ and $c_i= 3$ uniform across all agents. The opinion dynamics in this scenario, depicted in Fig.~\ref{fig:half}, reveal how initial polarization influences the pathway to consensus. Notably, the system exhibits a persistent polarization, which is not converged even at large imaginary times while differences among closest neighborhoods in middle positions become samller. This highlights the significant role of initial conditions in shaping long-term opinion distributions within the quantum framework.

\section{Consensus induced by leader}
\label{sec:leader}
The \textit{leader-follower} model introduces a leader agent whose opinion strongly influences all other agents. We examine the magnetization dynamics under varying leader influence strengths $d_i \equiv d$. 

%%%%%%%%%%%%%%%%%%%%%%%%%%%%%%%%%%%%%%%%%%%%%%%%%%%%%%
\begin{figure}[htbp!]
    \centering
    \includegraphics[width=0.95\linewidth]{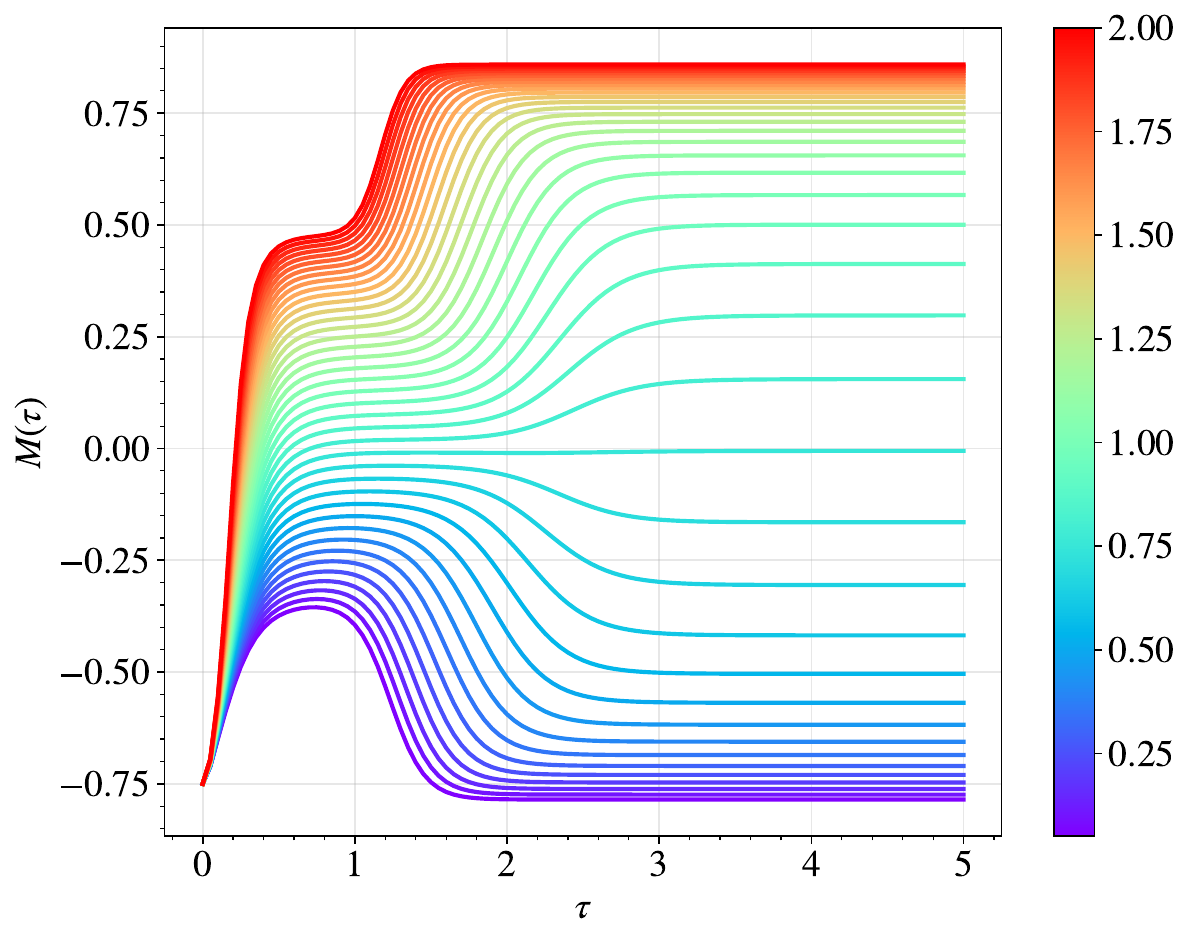}
    \caption{Magnetization dynamics $M(\tau)$ under varying leader influence strengths $d$. Different colored curves correspond to different values of $d$, increasing from weak (cool colors) to strong (warm colors).}
    \label{fig:leader-mag}
\end{figure}
%%%%%%%%%%%%%%%%%%%%%%%%%%%%%%%%%%%%%%%%%%%%%%%%%%%%%%
As shown in Fig.~\ref{fig:leader-mag}, increasing $d$ systematically drives the system toward a different consensus state. In particular, the original consensus plateau at $\tau \sim 2$ shifts to a higher magnetization as $d$ increases, indicating that the leader effectively aligns the opinions of other agents with its own. This behavior reflects a competition between the leader’s influence and the agents’ initial beliefs, resulting in a transition from a low-magnetization state to a high-magnetization state once the leader strength exceeds a critical value, $d \simeq 0.75$. Around this threshold, the metastable state coincides with the final consensus. Nevertheless, this behavior alone does not fully characterize the collective dynamics, as the network connectivity plays a crucial role in determining the system’s response to leadership.

\section{Hardware implementation details}\label{app:Hardware implementation details}
%%%%%%%%%%%%%%%%%%%%%%%%%%%%%%%%%%%%%%%%
\begin{figure}[htbp!]
    \centering
    \includegraphics[width=0.95\linewidth]{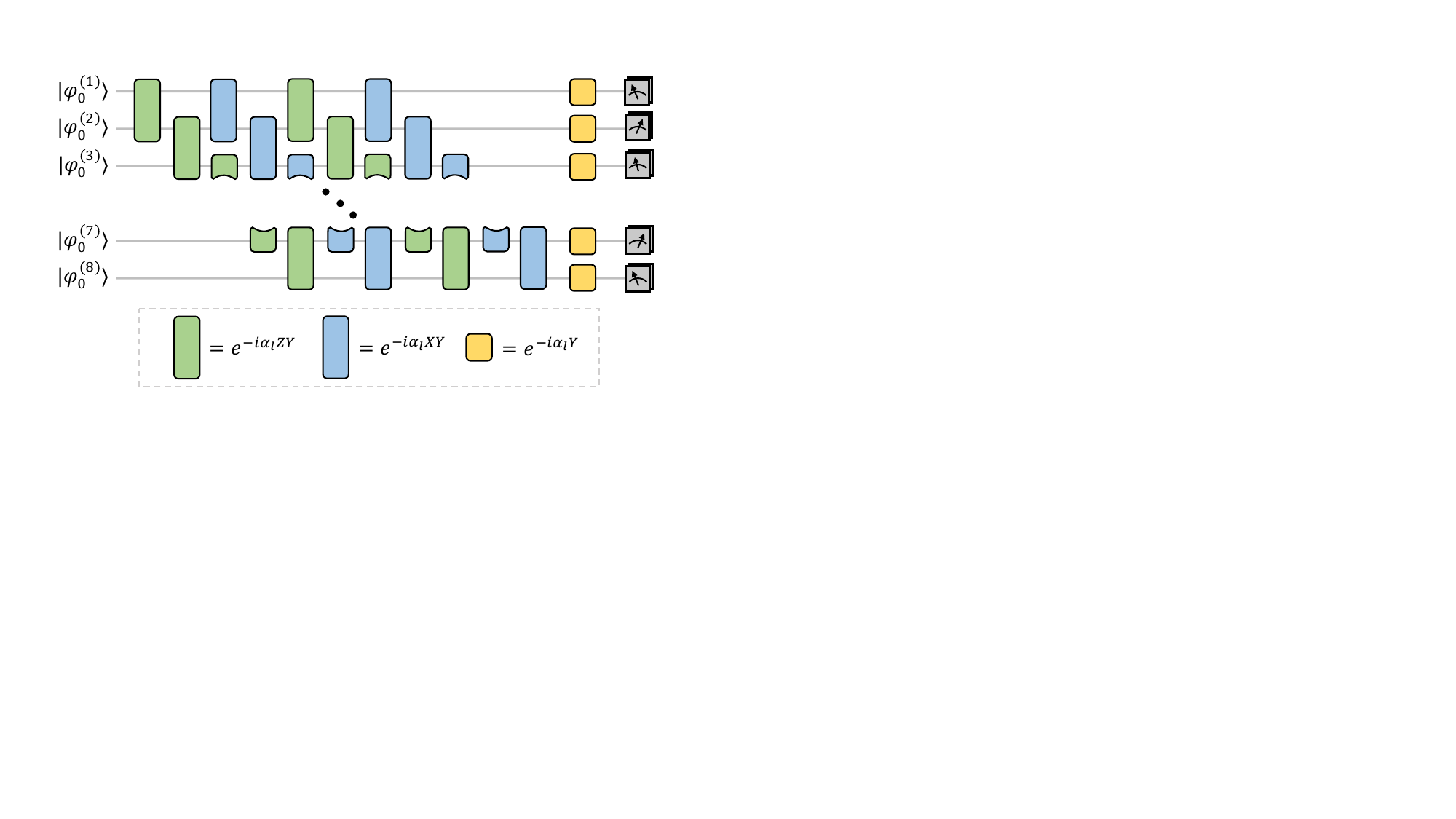}
    \cprotect\caption{Imaginary Hamiltonian variational ansatz to perform QITE. The initial state $\ket{\varphi_0^{(i)}}$ is prepared by $R_Y$ rotations. The green, blue, and yellow gates are $R_{ZY}$, $R_{XY}$, and $R_{Y}$ rotations with all-free parameters.}
    \label{fig:ansatz}
\end{figure}
%%%%%%%%%%%%%%%%%%%%%%%%%%%%%%%%%%%%%%%%
The imaginary-time dynamics on quantum hardware are performed using the variational quantum imaginary-time evolution (VarQITE) algorithm~\cite{McArdle_19}. This algorithm performs QITE on a parametrized quantum circuit $\ket{\phi[\bos{\alpha}(\tau)]}=U(\bos{\alpha})\ket{\varphi_0}$, where $U(\bos{\alpha})=U_L(\alpha_L)\ldots U_1(\alpha_1)$ is a series of parametrized unitary quantum gates. Then, QITE state $\ket{\varphi(\tau)}$ in Eq.~\eqref{eq:QITE} of the main text is approximated using $\ket{\phi[\bos{\alpha}(\tau)]}$ by minimizing their McLachlan distance using variational principle. This variational principle gives us the time derivative of the variational parameters $\ud \bos{ \alpha}/\ud\tau|_{\tau_0}$ at a given imaginary time $\tau_0$. Then the variational parameters at the next time slice $\tau_0+\delta \tau$ are given according to the Euler method
\begin{align}
    \bos{\alpha}(\tau_0+\delta \tau)\simeq \bos{\alpha}(\tau_0)+\left.\frac{\ud \bos{ \alpha}}{\ud\tau} \right|_{\tau_0}\times \delta\tau.
\end{align}
Using this formula iteratively, we derive the Euler evolution of free parameters $\bos{\alpha}(\tau)$ at any $\tau>0$.

For the opinion network on a one dimensional chain with $8$ agents, the parametrized quantum circuit used in our hardware demonstration is illustrated in Fig.~\ref{fig:ansatz}. Since the initial state $\ket{\varphi_0}=\prod_{i=1}^8\otimes \ket{\varphi_0^{(i)}}$ is the unique ground state of $H_0$, it is prepared by the single-qubit $R_Y$ rotations
\begin{align}
    \ket{\varphi_0^{(i)}}=\sin\frac{\theta_i}{2}\ket{0}-\cos\frac{\theta_i}{2}\ket{1}=e^{i(\pi-\theta_i)Y/2}\ket{0}
\end{align}
with $H_0$ parameters $\{\theta_1=(1-10^{-3})\pi,\theta_{1<i\leq 8}=0\}$. Then, we use the imaginary Hamiltonian variational ansatz~\cite{PhysRevA.111.032612}  composed of a sequential of $R_{\sigma_l}(\alpha_l)=e^{-i\alpha_l \sigma_l}$ rotations with $\sigma_l\in\{ZY,XY,Y\}$. These $\sigma_l$s are selected due to the time-reversal symmetry possessed by the full Hamiltonian $H$, and we set $\alpha$ as free variational parameters for each $R_{\sigma_l}(\alpha_l)$. These rotation gates can be decomposed into CNOT and single-qubit gates, and realized on the superconducting quantum chip. Thus, $L=36$ free parameters $\{\alpha_l\}$ in Fig.~\ref{fig:ansatz} are evolved according to the McLachlan variational principle. In the hardware implementation, the Euler evolution of $\bos{\alpha}(\tau)$ is determined by the classical statevector simulator with $\delta \tau=0.02$, and the opinion dynamics $\bra{\phi[\bos{\alpha}]}Z_i\ket{\phi[\bos{\alpha}]}$ is measured on the hardware.

To mitigate the coherent and incoherent noise during the execution of the quantum circuits, dynamic decoupling~\cite{Ezzel2023} and randomized compiling~\cite{PhysRevX.11.041039} are used to suppress the qubit decoherence and the coherent noise of two-qubit gates. The incoherent noise is mitigated by assuming a global depolarizing channel in the finally derived quantum state~\cite{Urbanek_21,Vovrosh_21}.

\end{document}